\documentstyle[12pt]{article}
\input math_macros
\input my_mac

\def\frac#1#2{{\textstyle {#1 \over #2}}}

\def\Eq{\begin{equation}}   \def\Endeq#1{\label{#1} \end{equation}}
\def\Eqa{\begin{eqnarray}}  \def\Endeqa#1{\label{#1} \end{eqnarray}}
\def\M{{\cal M}}

\def\gev{\rm\,GeV}  \def\tev{\rm\,TeV}

\def\alphaket{| \alpha \rangle}
\def\aeket{| \alpha_E \rangle}

\def\O{ {\cal O} }
\def\GB{ \Gamma_{\beta} }
\def\GH{ \Gamma_{\hbar} }

\begin{document}
\begin{titlepage}

\begin{center}
January, 1992      \hfill       HUTP-92-A009\\
\vskip .5 in
{\large \bf On Tunneling at Finite Energies and Temperatures}
\vskip .3 in
{
  {\bf Stephen D.H. Hsu}\footnote{Junior Fellow, Harvard Society of
     Fellows. Email: \tt Hsu@HUHEPL.bitnet}
   \vskip 0.3 cm
   {\it Lyman Laboratory of Physics,
        Harvard University,
        Cambridge, MA 02138}\\ }
  \vskip 0.3 cm
\end{center}

\vskip .5 in
\begin{abstract}
The rate of vacuum transitions for a system at finite temperature has
contributions from both classical thermal fluctuations and quantum tunneling.
The first is given in terms of the free energy of a sphaleron configuration
by $\GB \sim exp(- \beta F_{sphaleron})$, while the latter,
$\GH$,
is a complicated function of finite energy tunneling rates.
It is usually assumed that $\GH$ is always negligible at temperatures
greater than of order the mass of the lowest excitation.
We show that models exist in which, even at large temperatures, $\GH >> \GB$.
We examine the issue in the (1+1) Abelian Higgs model,
as well as in the case of electroweak (B+L) violating processes.
We show how the persistence of the cosmological baryon asymmetry
yields a bound on the inclusive two particle cross section
$\sigma_{2 \rightarrow All} (E)$ as a function of center of mass energy.

\end{abstract}
\end{titlepage}

\renewcommand{\thepage}{\arabic{page}}
\setcounter{page}{1}
\section{Quantum and Thermal Tunneling}

Consider a quantum system which exhibits a finite energy barrier separating two
local minima. An example in one-dimensional qauntum mechanics is the
double well potential. In field theory the barrier exists in
configuration space, with a saddlepoint at the lowest point on the barrier.
In the electroweak theory, the saddlepoint is the now famous
sphaleron\footnote{In this letter we will use the word sphaleron to refer
to any unstable saddlepoint configuration.}
configuration of
Klinkhammer and Manton \cite{sphal}, while in the case of a metastable false
vacuum it is an O(3) slice of Coleman's bounce \cite{vt,Coleman}.

Now consider a thermal distribution of excitations at temperature T
of one of the local vacua. It is an important problem to determine the rate
$\Gamma$ of vacuum transitions given such an initial distribution. There are
two contributions to this rate:
\begin{equation}
\Gamma(T) = \GB(T) + \GH(T).
\end{equation}
The first term on the right corresponds to the rate of classical,
thermal fluctuations
{\it over} the barrier. Affleck, Linde \cite{ImF,Linde}
and others have shown that this rate
can be computed semiclassically
in terms of the free energy of the saddlepoint configuration,
times a determinant of the quadratic fluctuation operator $O$,
\begin{equation}
\GB = [Det~  O  ]^{-\frac{1}{2}}~ exp( - \beta F_s )
\end{equation}
An intuitive picture of
this result is that $exp( - \beta F_s )$
 represents the probability of finding a
configuration in the thermal distribution which is at the saddlepoint, ready
to pass over.
It is worth noting that $\GB \rightarrow 0$ as $T \rightarrow 0$,
and that $\GB$ neglects quantum tunneling effects.

$\GH$ is defined as the contribution to the transition rate from
quantum tunneling. Each energy eigenstate of the thermal distribution has
some (non-perturbative) probability of tunneling through the barrier in a
classically forbidden manner. Therefore,
\begin{equation}
\GH = \frac{1}{Z} \sum_{\alpha} exp(-\beta F[\alpha])~ \sigma(\alpha),
\end{equation}
where $Z = \sum exp(- \beta F)$ is the partition function, $F[\alpha]$ is
the free energy of the configuration $\alphaket$, and $\sigma(\alpha)$ is the
probability of tunneling through the barrier from an initial state $\alphaket$.
$\GH$ is a purely quantum process, and in the limit of zero temperature is
just the rate of zero energy vacuum tunneling given by the usual instanton
calculation \cite{t'Hooft,Coleman}.

Since $\GB$ is zero at zero temperature,
it is clear that $\GH$ must saturate $\Gamma$ at sufficiently low
temperature. However, it is usually assumed that at any reasonable
temperature
(i.e. $T \sim \mu$, where $\mu$ is the mass of the lowest excitation)
the classical thermal rate $\GB >> \GH$. This argument was first presented
by Affleck \cite{ImF}, in a quantum mechanical WKB analysis.
Affleck found a narrow crossover region between $\GH$ and $\GB$ at temperature
$T_0 = \omega / 2 \pi$. Here $\omega^2$ is the curvature at the saddlepoint,
which is treated in a parabolic approximation.

Using results from finite energy tunneling in field theory,
we find that $\GH$ is not exponentially smaller than $\GB$ even at
temperatures $T >> T_0$. This is in agreement with the model considered by
Affleck, where the enhancement of $\GB$ over $\GH$ at high temperature occurs
in the prefactor.
However, we shall see that there exist models in which exactly
the opposite of the common wisdom is true: $\GH >> \GB$ for a
large range of temperature. This behavior is connected with
tunneling probabilities $\sigma (\alpha)$ which grow rapidly with
energy. In the model considered by Affleck, the shape of the barrier is
assumed to be such
that tunneling is dominated by configurations very close to, with
energies just less than, the sphaleron. This leads to the coincidence that
the imaginary part of the free energy, Im[F], gotten by summing over
imaginary parts for individual states in the thermal ensemble happens to
agree to exponential accuracy with $\GB$, which is gotten from the
sphaleron itself. Below we will consider models for which this is not the
case, and contributions to Im[F] from states with $E << E_s$ are important.

It is clear that the quantities $\GB$ and $\GH$ are distinct, as can be
demonstrated by the following argument. $\GB$ depends only on the `height' of
the energy barrier, and the small fluctuations about that point. On the other
hand, the individual rates $\sigma (\alpha)$ depend on the `width'
of the energy barrier. These statements are easily made quantitative in one
dimensional quantum mechanics, as $\sigma (\alpha)$ can be computed in terms
of WKB integrals $\int dx \sqrt{2(E_{\alpha} - V(x))}$ between turning points.
One
can now vary $\GH$ while keeping $\GB$ fixed by adjusting the `width' of the
barrier while keeping its `height' and local curvature fixed.
One possible result of such adjustments is a delta function-like potential
for which tunneling is enhanced and classical transitions suppressed.
These arguments
can be generalized to field theory if one now refers to the (infinite
dimensional) barrier in configuration space instead of the quantum mechanical
potential.
{}From the above arguments alone it seems plausible that there could exist
models in which $\GH > \GB$, even for large temperatures. One merely has to
consider a `narrow' barrier which readily allows quantum tunneling (see
figure 1).

The computation of $\sigma (\alpha)$ in field theory is a difficult
problem that has been considered previously by many authors \cite{MinkT}.
However, a recent result by Rubakov and collaborators (KRT) \cite{KRT}
allows a computation of
$\sigma (\alpha)$ for certain special initial states which correspond to `most
likely escape paths' (in the terminology of the authors of references
\cite{MinkT}) at fixed energy. More specifically, they find the most likely
inital state for tunneling, as well its tunneling probability
$\sigma(\alpha_E)$, among all states with some fixed energy E.

KRT find that tunneling at fixed energy E is dominated by periodic instanton
solutions \cite{KRT}, whose action gives the exponential factor in the
transition rate. The most probable tunneling state, which we denote $\aeket$,
is the analytic
continuation to Minkowski spacetime of the periodic instanton solution at its
turning point, where $\dot{\phi} (x,t) = 0$. In other words, the total
transition probability from all states of energy E is dominated by the
contribution from the most probable initial state corresponding to the
periodic instanton:
\begin{equation}
\sum_{\alpha} \delta ( E_{\alpha} - E ) \sigma (\alpha)
        \simeq \sigma (\alpha_E) ~\sim~ exp( - S_E ),
\end{equation}
where $S_E$ is the action of the periodic instanton. $S_E$ at $E=0$ coincides
with the standard instanton action, $S_0$, while $S_E \rightarrow 0$ as $E
\rightarrow E_{s}$. We can parametrize $S_E$ in the following way:
\begin{equation}
S_E \equiv S_0 (1 - P(E/E_{s}) ),
\end{equation}
where $P(x)$ satisfies $P(0) = 0$ and $P(1) = 1$.
It is worth mentioning that although the KRT periodic instantons give us a
concrete method for computing the function $P(x)$ and finding the states
$\aeket$, the form of the above equations follows merely from the assumption
of the existence of a most probable tunneling state.

We now note that the expression (3) for $\GH$ can be rewritten
\begin{equation}
\GH \simeq \frac{1}{Z} \sum_{E} exp(-\beta F[\alpha_E])~\sigma(\alpha_E)
\end{equation}
This in turn can be rewritten as an integral over energy if we can find a
simple relation between the free energy F of the configuration $\aeket$ and
E itself. The free energy of a given field configuration can be computed from
the finite temperature effective action at temperature T \cite{us}.
F is therefore
temperature dependent itself, and differs from E for a given configuration by
the effects of thermal loop corrections to the effective potential and higher
derivative terms. The leading thermal loop corrections yield corrections to
the zero temperature energy functional of order $\O (\lambda T^2 / m^2)$,
where $\lambda$ is a small coupling and $m$ is the mass of a single particle
excitation. Therefore, there is always a temperature regime,
$ m^2 << T^2 << m^2 / \lambda $, in which the energy functional and free
energy functional differ only by a small amount, or
$ F[ \alpha ] \simeq E_{\alpha}$.
In that temperature range we have
\begin{eqnarray}
\GH & \propto & \int dE~ exp( - \beta F[ \alpha] ) ~~\sigma(\alpha_E) \\
    & \simeq & \int_0^{E_{s}} dE~
             exp( - \beta E - S_0 (1 - P(E/E_{s} ) )) \\
    & \propto & e^{-S_0} \int_0^1 dx~ exp[ - (\beta E_{s}) x
                                       + S_0 P(x)    ].
\label{GH}
\end{eqnarray}

In the above expressions we have not retained dimensionful prefactors. This
is because in practice it is extremely difficult to compute the determinental
prefactors that multiply the exponential factors in $\GB$ or $\GH$.
(Dimensional analysis suggests that they are both of order $T^d$,
where d is the
dimension of spacetime in which we are working). In this
letter we will content ourselves with studying the exponential behavior of
these functions, which should give the dominant behavior in any regime where
couplings are small and the semiclassical approximations are valid.

Before proceding to study $\GB$ and $\GH$ in specific models, let us first
perform a simple calculation. Let $P(x) = x$, and take T to be a temperature
for which $\beta F_s << S_0$. In the models to be considered below, the
latter approximation is true at temperature greater than or of order the
mass of the lowest particle state (e.g. in the electroweak theory with
$M_{Higgs} \simeq M_W$, $S_0$ is of
order a few hundred, while $\beta F_s$ approaches zero at $T \sim M_W$).
With these approximations, the integral expression for $\GH$ can be evaluated
directly, giving
\begin{eqnarray}
\GH &\propto& e^{-S_0} \int_0^1 dx~ exp( x ( S_0 - \beta E_s) \\
    &\simeq& e^{-\beta E_s} / S_0.
\end{eqnarray}
Therefore, for this choice of $P(x)$ we find that $\GH$ and $\GB$ are
exponentially of the same order for a large range of temperatures. $\GH$
is suppressed by a factor of $1/S_0$, however this factor is not meaningful
given the uncertainties in the non-exponential prefactors.

A choice of $P(x)$ for which the finite energy tunneling grows even faster,
e.g. $P(x) = x^a$, where $a < 1$, would yield an even larger ratio of
$\GH / \GB$. Taking the limit $a \rightarrow 0$ would correspond to a theory
in which finite energy tunneling becomes important at very low energy, and in
which $\GH >> \GB$ for all temperatures. This would correspond to a barrier
which was extremely `narrow' relative to its `height'.
On the other hand for a theory with $a >> 1$, which corresponds to strong
suppression of finite energy tunneling, we can still derive a lower bound on
$\GH$ which is only down by a factor $1/a$. One arrives at this bound by
replacing the function $P(x) = x^a$ by a linearly increasing function which
is zero until $x = 1 - 1/a$, and increases to one with slope a.

In the following section, we will study a (1+1) dimensional
model in which it is possible to compute $P(x)$ to leading order in a small
parameter by use of KRT periodic instantons. We will find that the
rate for finite energy tunneling grows even faster than for $P(x) = x$, and
hence that the ratio $\GH / \GB$ is even larger than in the example studied
above.

In the last section we will use the fact that the inclusive-inclusive
(B+L) violating cross section is an upper bound for the two particle-inclusive
cross section to arrive at a cosmological bound on the
two body cross section.

\renewcommand{\thepage}{\arabic{page}}
\section{A (1+1) model with $\GH >> \GB$}

It is clear from previous arguments that one can construct
quantum mechanical potentials with the property that
tunneling is enhanced relative to thermal `hopping' over the barrier
(see figure 1).
One might, however, find these systems contrived, and wonder whether the
same type of behavior can arise in a field theory, where the shape of the
energy barrier is determined dynamically by the interactions themselves.
In this section we point out that the (1+1) Abelian Higgs
model is
an example of a field theory with the desired properties \footnote{This
was noticed independently by Sergei Khlebnikov, whom the author thanks
for conversations.}.

The Abelian Higgs model is nothing more than scalar electrodyamics with
spontaneous symmetry breaking. The Euclidean action for this model is
\begin{equation}
S = \int dx^2~ [ \frac{1}{4} F^2_{\mu \nu} +
   | (\partial_{\mu} - ig  A_{\mu}) \phi |^2
    + \lambda ( | \phi |^2 - v^2  )^2  ]
\end{equation}
This model posesses instanton solutions \cite{RU} which are equivalent to
Abrikosov-Nielsen-Oleson vortices \cite{vortices}. Let us consider the
limit $\lambda >> g^2$, which corresponds to $M_{\phi} >> M_A$, where
$M_{\phi} = 2 \sqrt{\lambda} v$ is the scalar mass and $\M_A = \sqrt{2} gv$
is the gauge boson mass. In this limit the vortex (instanton) consists of a
small core of size $M_{\phi}^{-1}$ in which $\phi \rightarrow 0$. Outside
the core, the scalar field has vacuum expectation value $<\phi> = v e^{i
\theta(x)}$, with $\theta(x)$ exhibiting at least one non-trivial winding
as $x$ goes around the core. The gauge fields are chosen to cancel the
gradient energy in the phase, so that the vortex (instanton) has finite
energy (action).

The existence of a finite action instanton solution corresponds to a finite
potential barrier separating different vacua. In this case, the height of
the barrier (or energy of the sphaleron) is given by $E_{s} \simeq
v^2 M_{\phi}$ \cite{AHS}. KRT \cite{KRT} have constructed the periodic
instantons for this theory by perturbing about the zero energy instanton in
a low-energy expansion. They find
\begin{equation}
\sigma(E) \sim exp( 4 \pi v^2 ln ( \frac{(E + M_A)}{ E_s} )   ),
\end{equation}
up to corrections that are small for $E/E_s << 3 \pi$.
At $E=0$, this corresponds to twice the zero energy instanton suppression.
However, the logarithmic behavior of the exponent implies that $\sigma(E)$
rises quickly with energy.

Using this result for $\sigma(E)$ we can rewrite
equation (\ref{GH}) as
\begin{equation}
\GH \sim \int_{0}^{1} dx~ exp( - \beta E_s x + 4 \pi v^2 ln (x+a) ),
\end{equation}
where $x = E/E_s$ and $a = M_A/E_s$. This integral can be computed in the
saddlepoint approximation, because it is dominated by the stationary point
of the exponent
(i.e.  $x_0 = \frac{4 \pi v^2}{ \beta E_s} - a
     \simeq \frac{4 \pi v^2} { \beta E_s} $ ).
This yields
\begin{equation}
\GH \sim exp[ - 4 \pi v^2 ( 1 + ln( \frac { \beta E_s }  {4\pi v^2} ) ) ].
\end{equation}
When we compare to $\GB \sim e^{-\beta E_s} = e^{ - v^2 \beta M_{\phi} }$, we
see that it easy to have $\GH >> \GB$, for temperatures in the range
$M_A << T << M_{\phi}$. One can speculate that the above results imply
that there exist tunneling
paths comprised mainly of gauge fields $A_{\mu}$, which
in this temperature range are more accessible
to the system than the sphaleron, which has a large Higgs
component.

\renewcommand{\thepage}{\arabic{page}}
\section{Electroweak (B+L) violation - a bound on the Holy Grail function}

We now turn to the more complicated phenomena of (B+L) violation in the
electroweak theory. Much attention has been paid to the
computation of $\GB$ in terms of the sphaleron free energy \cite{KRS,AM}.
The computation of $\GH$ is considerably more complicated, due to our
ignorance of the behavior of finite energy tunneling cross sections. We know
neither the most likely tunneling states nor the behavior of the cross
section,
except through a low energy expansion about the zero energy constrained
instanton \cite{KRT}.

An additional complication is that, due to the large
number of fields in the standard model (specifically, due to the 3 colors of
top quark, each with sizable Yukawa coupling to the Higgs), the free energy
of a given configuration $F[ \alpha]$ begins to differ considerably from its
energy $E_{\alpha}$ even at low temperatures. In fact,
if the top quark is very heavy, symmetry restoration
occurs at $T \sim M_{Higgs}$,
which can be as low as $48 \gev$ \cite{LEP}.
This makes the evaluation of equation (\ref{GH}) much more difficult.

In this section we will attempt a crude estimate for $\GH$. By requiring that
$\GH$ not wash out any existing baryon asymmetry
present after the electroweak phase transition, we will recover a bound on
the function $P(x)$, which describes inclusive-inclusive (B+L) violating
cross sections. In this analysis we will only obtain bounds on functions
$P(x)$ which describe rapid growth of the inclusive cross section with
energy. In other words, if the total rate $\Gamma$ is dominated by
$\Gamma_{\beta}$ just after the phase transition,
we obtain no information about $\GH$.
Since $\sigma_{All \rightarrow All} (E) > \sigma_{2
\rightarrow All} (E)$ for all E, we will therefore also obtain a bound on
the behavior of the second cross section. The function analogous to
$P(x)$ in the case of $\sigma_{2 \rightarrow All} (E)$ has been christened
the `Holy Grail' function, f(x). We therefore obtain an inequality
$P(x) > f(x)$.

First, let $H(x)$ denote the prefactor times integrand of equation
($\ref{GH}$),
except that here we cannot use the approximation $F [ \alpha ] \simeq
E_{\alpha}$, so we use the Free energy of the most probable tunneling path
with energy $E = E_s x$, which we denote $F[ \alpha_x]$:
\begin{equation}
H(x) = exp[ - \beta F[ \alpha_x ]  - S_0 ( 1 - P(x) ) ].
\end{equation}
Therefore, $\GH = \int_0^1 dx~ H(x)$, and we can use the following
inequality as a lower bound on $\GH$:
\begin{equation}
\GH ~>~ \int_{x_0}^1 dx~ H(x) ~>~ (1 - x_0 ) H(x_0).
\end{equation}
The above inequality is valid for any $x_0$ in the interval $[0,1]$,
provided that $H(x)$ is monotonically increasing. Since we will not use the
prefactor $(1 - x_0)$ in our bound, we are actually much less sensitive to
this asumption, and only require that $H(x)$ not exhibit large oscillations
over small changes in x.

Now, consider the early universe just after the electroweak phase
transition, at temperature T.
In order to preserve any baryon asymmetry present at that era,
we must require that $\Gamma = \GB + \GH$ be out of equilibrium (i.e. the
timescale for (B+L) violating transitions must be longer than the Horizon
time $\sim M_{Planck} / T^2$). While several authors \cite{washout}
have studied the
consequences of applying this condition to $\GB$,
here we assume that $\GB$ is
negligible and concentrate on $\GH$.
It is worth mentioning that
previous estimates of the washout rate are extremely
sensitive to the Higgs condensate $\phi_+$ immediately after the phase
transition. It has been shown that $\phi_+$ is strongly affected by
infrared corrections to the naive one-loop effective potential previously
used to study the transition \cite{us}. On the other hand,
the critical temperature $T_c$ is
relatively unchanged.
Below we will merely assume that $\phi_+$ is less than its zero temperature
value ($246 \gev$) by a factor of two or so. This is a reasonable
assumption, since if $\phi_+$ is large the nucleation rate of bubbles of
the new phase is strongly suppressed.

Let us recall the relevant parameters for the electroweak theory. Since we
are only able to make rough estimates in this calculation, we will not
insist on great accuracy here. The zero
energy instanton suppression is given by the t'Hooft factor \cite{t'Hooft}:
$S_0 = 16 \pi^2 / g^2 \simeq 350$. The sphaleron energy is roughly $10
\tev$ (there is only a weak dependence on the Higgs mass).
The critical temperature for the phase transition depends on both the top
and Higgs mass.
The lowest critical temperature (higher temperatures always
lead to larger $\Gamma$) is $T_c \simeq M_{Higgs}$ for very heavy top
quarks ($M_{top} \simeq 250 GeV$).
Finally, let the critical value of the washout rate, determined by the
expansion rate at T, be given by $e^{- S_c}$ ($S_c$ is roughly $50$).

Let us require that our lower bound for $\GH$ always be less than
$e^{-S_c}$. Taking the logarithm, and discarding the prefactor
$(1-x_0)$ which is only relevant for $x_0$ exponentially close to 1,
yields
\begin{equation}
\frac{ F[ \alpha_{x_0} ] }{ T } + S_0 ( 1 - P(x_0) ) > S_c.
\label{lb}
\end{equation}

It remains to estimate the Free energy
$F[ \alpha_{x_0} ]$. The state $| \alpha_{x_0} \rangle$ corresponds to some
multiparticle state with energy $E = E_s x$. It resembles a semiclssical
field configuration with contributions to its energy from kinetic and
potential energy terms in the energy functional. The leading thermal
correction by which the Free energy functional
differs from the energy functional can be accounted for in a shift of the
$SU(2)$
vacuum expectation value \cite{KRS}. However, the kinetic
energy (derivative) terms are probably only slightly
changed in the Free energy \cite{us}.
This being the case, and it being likely that the scalar vacuum expectation
value just after the phase transition
is significantly less than its zero temperature value, we
assume that the Free energy of a given configuration is decreased by an
amount roughly equal to its original potential energy, leaving only a
kinetic energy contribution. This is consistent with our view of the hot
plasma just after the phase transition - the correlation lengths are long
due to small effective masses for the gauge and Higgs fields.
Therefore, we have the following estimate:
$F[ \alpha_{x_0} ] \simeq E_s x_0 R$, where $R$ is roughly the fraction of
the energy of the configuration
$| \alpha_{x_0} \rangle$ which comes from kinetic energy.
We expect R to be roughly $.5$, although it could be significantly larger
or smaller.

Substituting the Free energy estimate into (\ref{lb}) yields
\begin{equation}
[ 1 - P(x_0) ] > \frac{S_c}{S_0} - \frac{ E_s x_0 R }{ S_0 T}.
\end{equation}
Substituting our preferred values for these parameters yields
\begin{equation}
[1 - P(x_0) ] >  1/7 ( 1 - \frac{200 \gev}{M_{Higgs}} R x_0 ).
\end{equation}
Finally, since $P(x) > f(x)$ for all $x$, we have the advertised bound on
the `Holy Grail' function:
\begin{equation}
f(x_0) < 6/7 + 1/7 ( \frac{200 \gev}{M_{Higgs}}~ R x_0 ).
\end{equation}
This bound is shown in figure 2, for $R = .5$. The dashed line corresponds
to $M_{Higgs} = 50 \gev$ and the solid line to $M_{Higgs} = 100 \gev$.

Note that for some values of the parameters above it appears that neither
$P(x)$ nor $f(x)$ can ever reach 1. Perhaps this should not concern us in
the case of $f(x)$, whose behavior is unknown, but we expect that $P(1)=1$.
Now, it is technically untrue that the bound can rule out 1 since
we have neglected the $ln (1 - x_0)$ term in the exponent, which becomes
important when $x_0$ is exponentially close to 1. But this escape leaves
possible only an extremely pathological behavior of $P(x)$.
The answer, of course, is that values of the parameters for which $P(x)=1$
is not allowed are already ruled out by requiring that $\GB$ not
wash out the baryon asymmetry.
For those values of the parameters, and under our assumptions about $T_c$
and $\phi_+$, washout is unavoidable due to $\GB$. Hence we obtain
no information about $\GH$ in that case.

The above arguments show that the existence of the
cosmological baryon asymmetry can yield bounds (however weak!)
on the behavior
of finite energy tunneling in both the two and multi-particle sectors.
In particular, the information we get is not confined to any low energy
expansion, and can constrain the behavior of cross sections at energies of
order the sphaleron energy.

\section{Conclusions}

In this letter we have addressed the contribution of finite energy
tunneling to the total rate for vacuum transitions in a system at finite
temperature. The assumption that $\GH$ is always negligible when
the temperature
is greater than of order the lowest particle mass has been shown to be
untrue. For a generic system, we expect $\GH \simeq \GB$
to exponential accuracy. We have presented examples of models in
quantum mechanics and field theory in which $\GH >> \GB$ for a
large range of temperature.

We have also derived a cosmological upper bound on the (B+L) violating cross
sections $\sigma_{2 \rightarrow All} (E)$ and $\sigma_{All \rightarrow
All} (E)$. This upper bound does not rely on a low energy expansion, and
applies even when E is of order the sphaleron energy.

\section{Acknowledgments}

The author would like to thank C. Glenn Boyd, Thomas Gould, Sergei Khlebnikov,
Richard Holman and Larry Widrow
for numerous discussions, and the UCSB Institute for Theoretical physics
for its hospitality while this work was conducted.
SDH acknowledges
support from the National Science Foundation under grant NSF-PHY-87-14654,
the state of Texas under grant TNRLC-RGFY106, and from the Harvard Society
of Fellows.

\end{document}